\documentclass[review]{elsarticle}

\usepackage{hyperref}

\usepackage{hhline}
\usepackage{amsmath,amssymb,amsfonts}
\usepackage{algorithmic}
\usepackage{graphicx}
\usepackage{textcomp}
\usepackage{enumerate}
\usepackage[linesnumbered,ruled]{algorithm2e}
\usepackage{caption}
\usepackage{xcolor}
\usepackage{amsmath}
\usepackage{booktabs}
\usepackage{graphicx}
\usepackage{stfloats}
\usepackage{tablefootnote}
\usepackage{subfigure}
\usepackage{amsmath}
\usepackage{balance}
\usepackage{float}
\usepackage[misc,geometry]{ifsym}
\usepackage{color}
\usepackage{url}
\usepackage{verbatim}
\usepackage{array}
\usepackage{graphicx}
\usepackage{xcolor}
\usepackage{array}
\usepackage{colortbl}
\usepackage{booktabs}
\usepackage{multirow}
\newcolumntype{P}[1]{>{\centering\arraybackslash}p{#1}}
\usepackage[utf8]{inputenc}
\usepackage{soul,xcolor}

\usepackage[utf8]{inputenc} 
\usepackage[T1]{fontenc}    
\usepackage{hyperref}       
\usepackage{url}            
\usepackage{booktabs}       
\usepackage{amsfonts}       
\usepackage{nicefrac}       
\usepackage{microtype}      
\usepackage{lipsum}		
\usepackage{graphicx}
\usepackage{natbib}
\usepackage{doi}

\journal{arXiv}

\begin{document}

\begin{frontmatter}

\title{Comparative Study of Machine Learning Test Case Prioritization for Continuous Integration Testing}

\author{Dusica Marijan}
\address{Simula Research Laboratory, Norway}

\cortext[mycorrespondingauthor]{Corresponding author}
\ead{dusica@simula.no}

\begin{abstract}
There is a growing body of research indicating the potential of machine learning to tackle complex software testing challenges. One such challenge pertains to continuous integration testing, which is highly time-constrained, and generates a large amount of data coming from iterative code commits and test runs. In such a setting, we can use plentiful test data for training machine learning predictors to identify test cases able to speed up the detection of regression bugs introduced during code integration. However, different machine learning models can have different fault prediction performance depending on the context and the parameters of continuous integration testing, for example variable time budget available for continuous integration cycles, or the size of test execution history used for learning to prioritize failing test cases. Existing studies on test case prioritization rarely study both of these factors, which are essential for the continuous integration practice. In this study we perform a comprehensive comparison of the fault prediction performance of machine learning approaches that have shown the best performance on test case prioritization tasks in the literature. We evaluate the accuracy of the classifiers in predicting fault-detecting tests for different values of the continuous integration time budget and with different length of test history used for training the classifiers. In evaluation, we use real-world industrial datasets from a continuous integration practice. The results show that different machine learning models have different performance for different size of test history used for model training and for different time budget available for test case execution. Our results imply that machine learning approaches for test prioritization in continuous integration testing should be carefully configured to achieve optimal performance.
\end{abstract}


\end{frontmatter}


\section{Introduction}

Continuous integration (CI) is an agile software development practice where software is realeased frequently following frequent code changes. Each change needs to be verified before a new change can be made and a new version of the code released. This process runs in CI cycles, also called builds or commits. Software testing is an integral step running iteratively and successively as part of continuous code integration. Each code integration is followed by an integration testing iteration, which is typically extensive, to prevent breaking a build. CI testing requires short turnaround between starting test execution and detecting faulty regressions, to enable fast feedback. This entails a short \textit{time budget} allocated to integration testing, which denotes the amount of time available for testing the code changes introduced in the latest commit. Short time budget requires testing in CI to be time-efficient \cite{niu, savor, parnin2017top, Marijan2019, 3183532, 8377636}. As a response to this challenge, researchers have proposed various test selection, minimization, and prioritization \cite{shi2019understanding, ali2020enhanced, Rothermel, 7272927, 7911878, 7928010, 7745984} approaches. In this work we specifically focus on \textit{Test Prioritization} (TP). TP consists in ordering test cases that are more effective in detecting faults to execute sooner. In this way, we ensure that the most important test cases are executed in a short time budget. However, in dynamic CI environments with frequent code changes, a time budget can vary across different CI cycles. Therefore, an efficient TP approach needs to adapt to varying time constraints across CI cycles. 
 
Furthermore, given that testing runs frequently in CI generating a large volume of test information, researchers have proposed \textit{history-based} TP approaches. These approaches use historical test execution information to speed up the detection of regression faults introduced by developers. \cite{Hem} suggests that history-based TP is an effective approach for rapid release software, while \cite{Srikanth} reports that using historical test failure information is a good indicator for test prioritization. However, history-based TP has its challenges. One common challenge is to decide how old historical information to use in TP. On the one hand, using too old history may capture old (irrelevant) failures which have been fixed and thus are not indicative of new failures. On the other hand, using too recent history may omit some relevant failures. To deal with this challenge, \cite{Elbaum} introduced the notion of time windows, to capture how recently tests were executed and failures exposed, which is further used for test selection in pre-submit and post-submit testing.  

\begin{figure*}
\centering
\includegraphics[trim=0.3cm 12cm 7.7cm 0cm, width=4.7in]{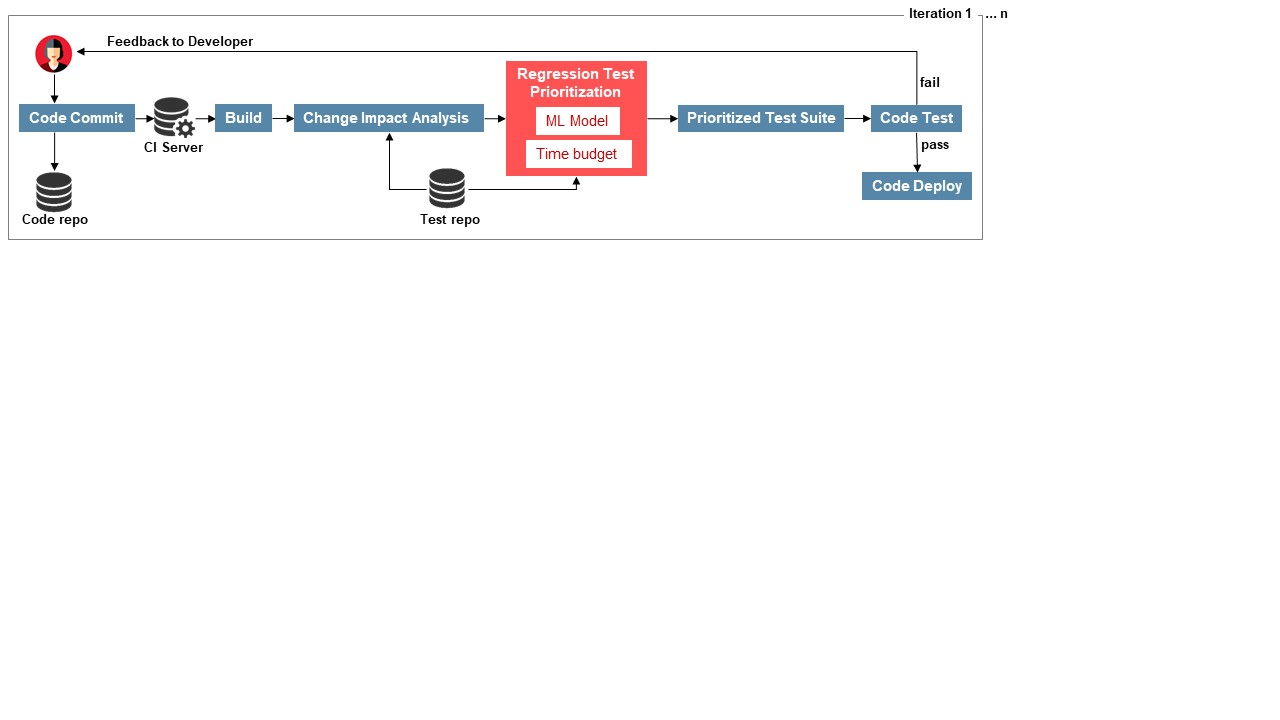}
\caption{An iteration of CI testing consists of a code commit, build, and test phase, before code deployment. Change impact analysis and regression test prioritization are used to speed up testing. Regression test prioritization can use test history to learn to prioritize test cases and ML to scale test prioritization process to the size of large test history.}
\label{fig_image1}
\end{figure*}

Now we illustrate one example of CI testing, describing daily practices and challenges of our industrial collaborator in the domain of testing configurable communication software in CI, shown in Figure~\ref{fig_image1}

Following a standard practice, code changes committed by developers are regression tested before they can be deployed to production. Several hundreds of changes made on a daily basis trigger the execution of several thousands of test cases. Change impact analysis is run to select the test cases impacted by the change. However, all impacted test cases cannot not fit the available time budget, and moreover, not all impacted test cases are equally useful in detecting faults. If test engineers were to manually select a subtest of tests produced by change impact analysis which they believe have the highest chance of detecting faults, such a process would be highly time-ineffeicient. Thus, test engineers have applied automated regression TP, as an established approach to improve the effectiveness of regression testing in CI. Specifically, given code changes and test execution history, the applied regression TP approach \cite{6676952} computes an ordered set of test cases that are impacted by the code changes, and that are of the highest historical fault detection ability. 
With this approach, test engineers can detect up to 30\% more regression faults compared to manual test selection guided by tester's expertize, for the same time budget. However, this approach is not well suited for processing a large set of historical test execution data. Following a recent research direction of using machine learning (ML) for  software testing, the goal of test engineers has been to develop a ML approach for TP that will be both time-efficient and have a high fault-detection efficiency as more test data becomes available.

While developing a ML-based test prioritization approach addressing the needs of our industrial partner, we observed that different ML models can have different fault-detection performance depending on the CI testing context. This was especially the case when we used ML models in CI cycles with different time budget (budget for testing) and when we used different size of test execution history for ML model learning. Therefore, we systematically studied how do these two parameters affect the fault-prediction performance of ML-based test prioritization models. 

In this paper we report the experimental results of the systematic comparison of four best-performing ML approaches reported in the literature on the task of test case fault-prediction: support-vector machines (SVM), artificial neural networks (ANN), gradient boosting decision trees (GBDT), and LambdaRank using NN. In addition, we compare the performance of ML-based approaches against two heuristic approaches: history-based TP approach $ROCKET$ \cite{6676952} and $Random$ test selection. We run the experiments on three industrial data sets from the CI practice: \textit{Cisco}, \textit{ABB}, \textit{Google}. 

In summary, our work makes the following contributions:
\begin{itemize}
\item Systematic analysis of how the size of test history affects fault-prediction effectiveness of learning-based test case prioritization.  

\item Systematic evaluation of the effectiveness of ML based test case prioritization approaches relative to variable time budget across CI cycles. 

\item Systematic comparison of the best-performing ML-based test case prioritization approaches reported in literature, evaluated on three industrial datasets. 
\end{itemize}

The paper is structured as follows. In Section 2, we review related work. In Section 3, we describe learning based test case prioritization and present the four ML-based test prioritization approaches evaluated in this study. Section 4 describes the experimental evaluation, while Section 5 presents the experimental results. We discuss the key findings of the study and conclude the paper in Section 6. 

\section{Related Work}
Recent studies have used ML for the problem of test case prioritization. Machalica uses a boosted decision tree approach to learn a classifier for predicting a probability of a test case failing based on code changes and  and subset of test cases \cite{8804462}. The approach was shown to reduces the testing cost by a factor of two, while ensuring that over 95\% of individual test failures and detected. Chen proposes another predictive test prioritization approach based on XGBoost \cite{3236053}. It studies test case distribution analysis evaluating the fault detection capability of actual regression testing. The approach has been used in practice, and has shown to significantly reduce testing cost. Motivated by the success of these two approaches based on gradient boosting, we selected the $GBDT$ as an evaluation candidate for our study. 

Busjaeger proposes a test case prioritization approach based on Support Vector Machine (SVM) \cite{2983954}, to learn a binary classifier to order test cases based on historical information. The approach has shown to outperform non-ML-based test case prioritization approaches in terms of fault-detection effectiveness. Lachmann \cite{Lachmann} uses SVM-Rank to prioritize test cases using test case failure information. The evaluation shows that SVM based approach to test case prioritization outperforms manual approaches by experts. Grano uses SVM and Random Forest (RF) to build a regression predictive model for assessing test branch coverage \cite{8368454} for the purpose of efficient test case generation for CI testing. The experimental results have shown good fault prediction accuracy of SVM for test case prioritization, therefore, we selected SVM as evaluation candidate in our study. 

Several test case prioritization approaches have been proposed using different forms of neural netowrks, such as Bayesian network \cite{4539555}, NN \cite{Mahdieh2020IncorporatingFE}, ANN \cite{Hosney}, RNN \cite{9023052}. Specifically, \cite{4539555} integrates a feedback mechanism and a change information gathering strategy to estimate the probability of a test case to find bugs. The approach has showed to enable early fault detection. \cite{Mahdieh2020IncorporatingFE} prioritizes test cases using a NN approach and the fault-proneness distribution of different code areas. The approach has showed to improve the effectiveness of coverage based test case prioritization. \cite{Hosney} uses the combination of test case complexity information and software modification information to train an ANN, to enable early fault detection. The approach has showed to improve fault detection effectiveness. \cite{9023052} proposes a gated recurrent unit trained on the time series throughput information to perform regression testing of web services. The results have shown good fault prediction performance. Following a good fault-prediction performance of these studies, we included the ANN approach in our evaluation study.

There are studies using reinforcement learning (RL) for test case prioritization, which focus on maximizing a reward when failing test cases are prioritized
higher \cite{Shi} or on using simpler ML models for RL policy design \cite{Rosenbauer}. Lima proposes a multi-armed bandit (MAB) approach to test case prioritization in continuous integration \cite{Lima}, which showed to outperform the RL approach in terms of fault-detection. However, we experimented with these approaches for test prioritization and they showed to be computationally expensive \cite{9609187}. Furthermore, Bertolino \cite{Bertolino} conducts an extensive experimental study  comparing RL against supervised learning for test case prioritization, and concludes that the RL approach is less efficient on this specific task. Because of our experience with RL and the experience reported by Bertolino, we did not select the RL and MAB approaches for our evaluation study, as our goal is to build a fast-running test case prioritization approach that can satisfy strict time constraints of short CI cycles. 

In the same study \cite{Bertolino}, Bertolino reports the best performing ML approach to test case prioritization in terms of fault-detection effectiveness are MART and LambdaMART. Motivated by this finding, we include LambdaRank in our evaluation study. LambdaRank is from the same family of learning to rank algorithms as MART, and we include it instead of MART, as MART is based on gradient boosted decision trees which we have already included in our study. 

\section{Learning Based Test Prioritization}
In CI development practices, testing is time-constrained and produces voluminous test history $H$, as CI cycles run fast and frequently. The test history $H$ contains test execution information for each CI cycle $C_i$, denoted as cycle history, where $i=1...n$ and $n$ is the number of CI cycles. Each cycle history consists of a test suite $T=\{T_1, T_2,...,T_n\}$ run in that cycle and the time budget of the cycle $B$. Each test suite $T$ contains the pass/fail execution status and execution time of each test case $t_i$. 

Given $H$, collected in runs in previous CI cycles, the goal of the learning-based test case prioritization is to predict which test cases will be effective in detecting faults in the current CI cycle $C_{n+1}$, ranked according to their probability of detecting faults. In addition to fault detection effectiveness, some approaches use test execution time $t$ as another prioritization criteria, which can be combined together \cite{6676952} to ensure that failing test cases are ordered higher, and among the failing test cases, those that execute faster are ordered higher.    

\begin{figure}[h]
\centering
\includegraphics[trim=0cm 12cm 15cm 0cm, width=4in]{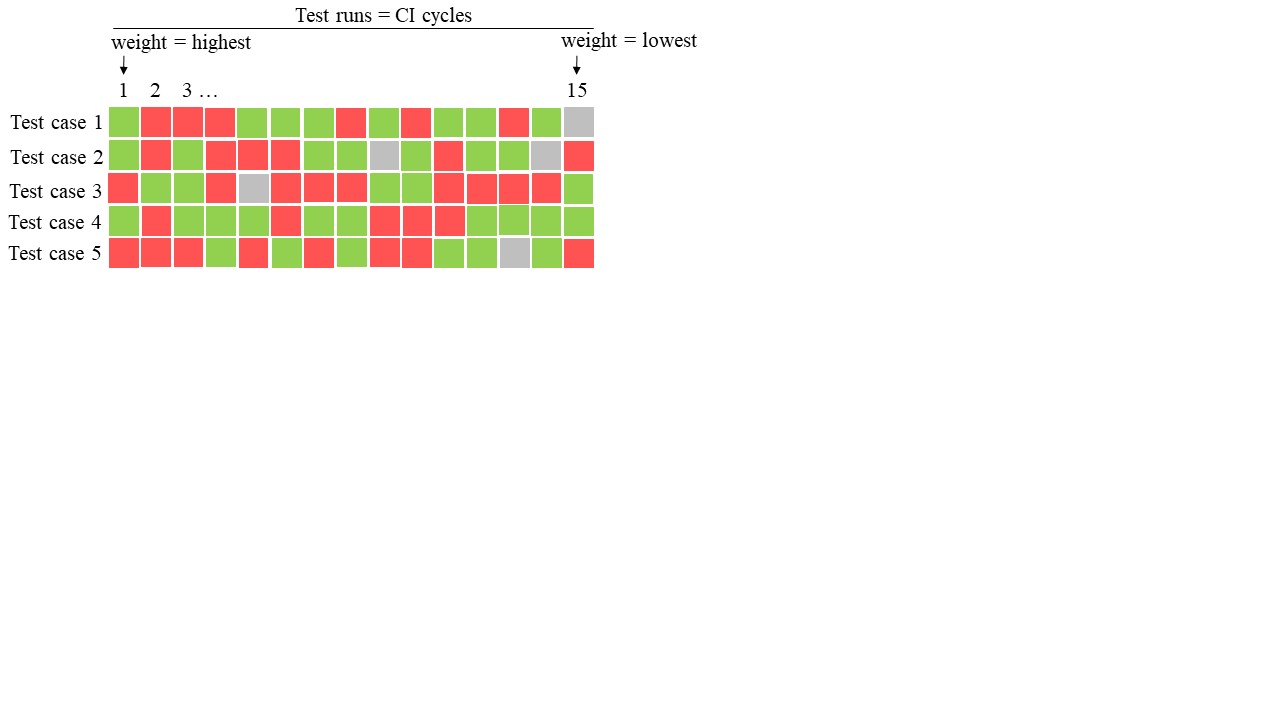}
\caption{Test history consisting of 15 CI cycles. Cycle 1 is the most recent and has the highest weight, while cycle 15 is the oldest and has the lowest wight. Test cases can change execution result between pass and fail in consecutive executions (CI cycles). Red: fail, Green: pass, Grey: inconclusive. In this work, we only deal with pass and fail test results.}
\label{fig_image2}
\end{figure}

In history-based test case prioritization, historical test failure records may be weighted, such that the highest failure weight corresponds to the failure exposed in the most recent test case execution and the failure in every precedent test execution is weighted lower. This ensures that the test cases that failed in the most recent run will be ordered higher (thus executed first), followed by a number of "older" failed test cases, depending on the available time budget $B$. Such "older" failed test cases are execution candidates as well, because tests can be flipping from fail to pass to fail again, as illustrated in Figure~\ref{fig_image2}. In case of ties, i.e. two or more test cases have the same failure probability, test cases should be ranked in the order of the shortest execution time $t$. We can define the problem of learning based regression test prioritization as follows: 

For a test case $T_i$ belonging to a regression test suite $T=\{T_1,T_2,...,T_n\}$, the goal of learning is to find a function $g:T \to C$, mapping the test case $T_i$ to a class $C_i$ (test rank) belonging to $C=\{C_1,C_2,...,C_m\}$, where $T_2$ is ranked higher than $T_1$ if $g(T_2) > g(T_1)$, $m$ is the number of test ranks. In binary classification $C \in \{0,1\}$. Each $T_i$ has its execution time $t_i$ and $n$ historical execution results $\{R_{i,1},R_{i,2},...R_{i,n}\}$, where $R \in \{0,1\}$ denotes a test pass or fail, and $n$ denotes the number of CI cycles (test executions). Although it is possible that the value of $t_i$ varies across different cycles, in this work we assume that $t_i$ is the average execution time of a test case across its CI cycles, as done in \cite{6676952}. 

\subsection{Selection of ML Approaches for Test Prioritization in CI}
As discussed in the related work, there are many ML approaches for test case prioritization. However, as we are interested in improving the efficiency of test prioritization in the CI practice, which is highly time-constrained, in our industrial case study we were looking for a time-efficient ML approach that can serve the need of generating prioritized test suites quickly. For example, we have previously experimented with  RL for test case prioritization in comparison with the NN approach on four industrial datasets \cite{9609187}, and have found the total runtime of the RL approach to be around 50 times higher than the runtime of the NN approach. This is consistent with the results reported by Bertolino \cite{Bertolino}. Therefore, we excluded RL approaches from this comparative study. Driven by the requirement to build a fast-running ML approach to test prioritization, we implemented four simpler types of classifiers for learning to prioritize regression tests, which have previously showed good fault detection performance, as discussed in the related work. The classifiers are learned on historical test execution results generated throughout several months of testing.   
For our evaluation study we selected the following ML classifiers: Support Vector Machine (SVM) classifier, Artificial Neural Network (ANN) classifier, Gradient Boosted Decision Tree (GBDT) classifier, and LambdaRank with NN (LRN) classifier.
 
\section{Experimental Evaluation}
The goal of the experimental study is to evaluate and compare the performance of four ML-based test case prioritization approaches discussed in Section 3 with the aim of answering the following research questions:
\begin{enumerate}[\bfseries RQ1]
\item How does the length of test execution history used for learning to prioritize test cases impact the fault-prediction performance of ML approaches? 

\item Which ML approach is more effective in predicting test cases with higher fault-detection effectiveness, for a given time budget, and how do they compare to heuristic-based test case prioritization approaches? 

\item Which ML approach is more time-efficient in a test prioritization task, and how do they compare to heuristic-based test case prioritization approaches? 
\end{enumerate}

\subsection{Experimental Dataset}
We perform experimental evaluation on three industrial datasets used for system-level testing in CI: \textit{Cisco}, \textit{ABB}, \textit{Google}. \textit{Cisco} dataset is used for testing video conferencing systems, provided by Cisco Systems. \textit{ABB} dataset \footnote{https://bitbucket.org/HelgeS/atcs-data/src/master/} is used for testing painting robot software, provided by ABB robotics. \textit{Google} dataset \footnote{https://code.google.com/archive/p/google-shared-dataset-of-test-suite-results/} is from a large scale continuous testing infrastructure provided by Google \cite{Google}. The datasets contain the information about the number of test cases, the number of test executions (CI cycles) for each test case and the historical fault-detection effectiveness of each test case in each execution as pass or fail. 
 
We summarize the datasets in Table~\ref{tab1}.

\begin{table}[h]
\centering
\renewcommand{\arraystretch}{0.9}
\caption{Evaluation datasets.}
\begin{tabular}{c | c  c  c }
\toprule
\textbf{Dataset} & \textbf{\# test cases} & \textbf{\# test executions} & \textbf{\% failed test cases} \\
\hline 
\textit{Cisco} & 550 & 6050 & 0.43\\
\textit{ABB} & 1488 & 149700 & 0.28 \\
\textit{Google} & 5507 & 12439910 & 0.01\\
\bottomrule
\label{tab1}
\end{tabular}
\end{table}

\subsection{Evaluation Baselines}
We compare the ML models for test prioritization one against the other, as well as against the automated TP approach $ROCKET$ \cite{6676952} that has previously shown to improve the effectiveness of manual practice of test selection at Cisco, and the $Random$ approach. 

$ROCKET$ prioritizes a set of test cases in CI testing based on historical test execution status and test execution duration. The basic principle of $ROCKET$ is that given the statuses of test cases' previous runs in successive CI cycles and their average execution time, the algorithm computes a priority value for each test case such to maximize early fault detection. More information about $ROCKET$ can be found here \cite{6676952}. We varied the length of historical information used for prioritization by $ROCKET$ from the most recent 20\% to the whole test history size available, with an increment of 20\%. Variable length of test execution history is not applicable to $Random$ heuristic, because it orders test cases randomly, without considering their historical fault-detection effectiveness during test selection.

\subsection{Evaluation Metrics}
We perform the comparison in terms of the following metrics: 

\textbf{\textit{APFD}} as the weighted average of the percentage of faults detected. 

\textbf{\textit{TDFT}} as the time to detect the first fault by a prioritized test suite. 

\textbf{\textit{TDLF}} as the time to detect the last fault by a prioritized test suite. 

\textbf{\textit{TRAIN}} as the training time of a ML model for test prioritization.

\textbf{\textit{PART}} as the running time of a prioritization algorithm, i.e. ranking time.
 
\subsection{Experimental Setup} 
First, for the evaluated ML-based approaches, for the purpose of model learning, we used a varying length of test history for all three datasets (\textit{Cisco}, \textit{ABB} and \textit{Google}), from the most recent 20\% (approximately corresponding to the most recent 20\% of CI cycles) to the whole test history available, with an increment of 20\%, which we denote as $H1$-$H5$. The basic idea of ML is that more data yields better performance. However, in the case of CI testing, using more historical cycles for learning may or may not mean better prediction performance \cite{7816510}, since some of the previous faults might have been fixed in previous CI cycles and in that case they are no longer good predictors of failing test cases. Next, we ran the 20 learned ML models for each of the three experimental test suites: \textit{Cisco}, \textit{ABB} and \textit{Google} to produce prioritized test suites. Next, we ran the two heuristic-based test prioritization approaches, $ROCKET$ and $Random$, for all three datasets. 

In the next part of the experiment, we selected the learned classifiers with the best size of test history used for model learning and produced the prioritized test suites. Next, we run the prioritized test suites to evaluate their fault-detection effectiveness using five varying values of the time budget ($B1$-$B5$). $B5$ corresponds to the average time required to run the whole test suite, and $B1$ corresponds to 20\% of that same budget. The remaining time budgets increase from $B1$ to $B5$ with increments of 20\%. By decreasing the time budget, we can assess the effectiveness of a test suite to detect failing test cases earlier, because a well performing ML predictor would prioritize failing test cases higher. 

In the final part of the experiment, we measured the time effectiveness of the TP approaches in terms of training time (for the ML approaches), ranking time, time to detect the first and last fault, and compared them with the heuristic-based TP approaches. We measure all the metrics on the \textit{Cisco}, \textit{ABB} and \textit{Google} datasets. 
  
Training the ML models requires parameter tuning. Specifically, to achieve good performance of the $GBDT$ model, we experimented with two hyperparameters, \textit{learning rate} and \textit{n\_estimators}. The learning rate affects the rate of adding new trees to the model. For example, a lower learning rate usually gives a more generalized learner. However, a lower learning rate needs more time for model training, and it requires a higher number of trees. Many trees may lead to overfitting. Therefore, choosing an optimal \textit{learning rate} and \textit{n\_estimators} is important for good performance of the $GBDT$ model. Similarly, the performance of the learned NN classifier is dependent on different parameters, such as the number of hidden layers and their sizes, activation function, and the number of epochs. To learn a well performing classifier, we performed an exhaustive hyperparameter tuning. We trained several classification models, while varying the number of hidden layers, and layer sizes for each layer. ReLu was used as the activation function for the hidden layers. Each network had 50 epochs, the training process of each network was iterated ten times, while measuring the average Mean Square Error (MSE) and Standard Deviation (SD) of MSE for all five networks.  
Finally, we chose the best performing target 3-layer network with the minimal MSE and SD.

\section{Results and Analysis}
In this section, we first analyse the experimental results answering the research questions, and discuss main threats to the validity of the reported results.

\subsection{RQ1: Effect of Test History Size on Fault-prediction Performance}
\begin{figure}[htbp]
\centerline{\includegraphics[width=6.5in]{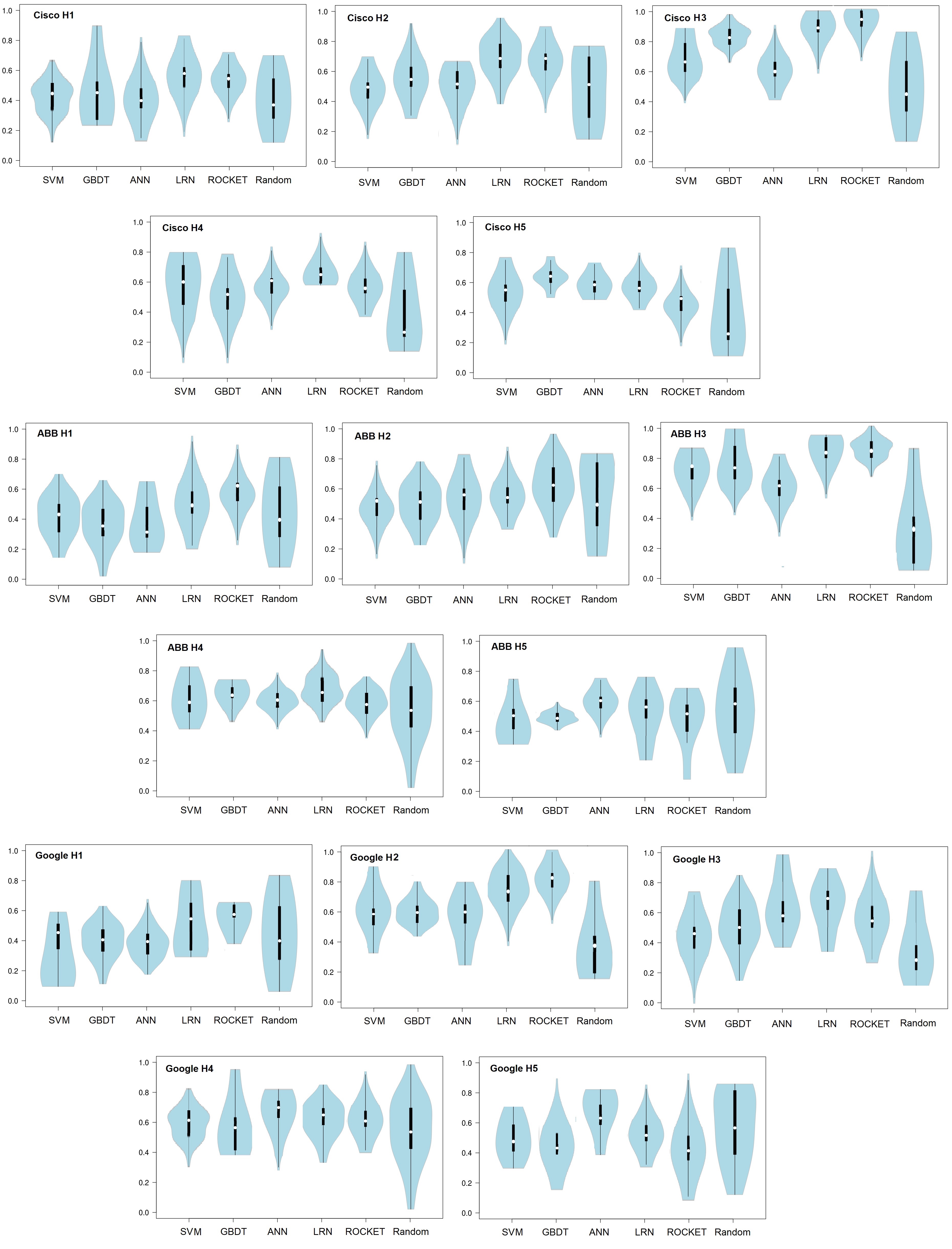}}
\caption{Performance of TP approaches in terms of APFD for different size of test history used for learning to prioritize ($H1$-$H5$) across three datasets: \textit{Cisco}, \textit{ABB}, and \textit{Google}.}
\label{fig_image 12}
\end{figure}

We show the fault-detection effectiveness of history-based TP approaches for different lengths of test history used for learning to prioritize in Figure~\ref{fig_image 12}. Overall, our experimental results indicate that the fault-prediction performance of the history-based approaches for TP (both ML based and $ROCKET$) varies depending on how much test execution history is used in learning to prioritize. Specifically, for the shortest length of test history ($H1$), all TP approaches achieve low performance. As the length of test history increases ($H2$), the fault-prediction performance of all TP approaches increases across all datasets. Increasing the length of test history further ($H3$) has a positive effect on all TP approaches for the \textit{Cisco} and \textit{ABB} datasets. However, for the \textit{Google} dataset we see a decrease in the performance for all TP approaches except $ANN$ in $H3$. Increasing the length of test history further ($H4$) has a negative effect on all TP approaches across all datasets except $ANN$ for the \textit{Google} dataset. Overall, we see that there is a less negative effect on $ANN$ compared to other ML approaches approaches. However, the results show that $ROCKET$ is more negatively affected by using older test history than the ML approaches. As we continue to increase the length of test history ($H5$) the performance of all approaches decreases, and more significantly for $ROCKET$ than for the ML approaches. Among the ML approaches specifically, we see that the $ANN$ approach is the less sensitive to using older test history than other ML approaches. Also the $ANN$ approach showed to be the most sensitive to using younger test history (for example, it has the worst performance out of all ML approaches for $H3$ on the \textit{Cisco} and \textit{ABB} datasets. 

In summary, the results indicate that the size of test execution history used for learning to prioritize affects the fault-prediction performance of TP approaches. For the \textit{Cisco} and \textit{ABB} experimental datasets, the optimal size of test history has shown to be $H3$, which corresponds to 60\% of test history. For the  \textit{Google} datasets, the optimal size of test history in our experiment was 40\%. This implies that the optimal size of test execution history decreases with the increase of test cycles. For example, the \textit{Google} datasets has longer test history of 2259 cycles, while the \textit{Cisco} and \textit{ABB} datasets have only around 100 cycles. This may also mean that the optimal size of test history is dependent on the frequency of bug fixing and code commits, i.e. the frequency of CI cycles. For datasets with longer test history less percentage of it should be used for test prioritization compared to the datasets with shorter history. 

\subsection{RQ2: Fault-detection Effectiveness for Different Time Budget}
\begin{figure}[htbp]
\centerline{\includegraphics[width=6.5in]{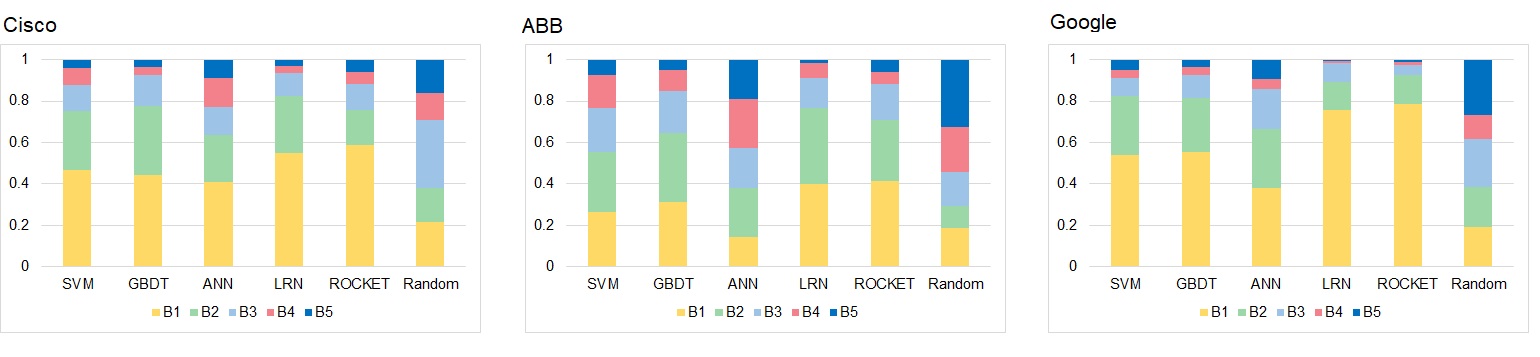}}
\caption{Performance of TP approaches in terms of average APFD for different size of test budget ($B1-B5$) across three datasets: \textit{Cisco}, \textit{ABB}, \textit{Google}.}
\label{fig_image 13}
\end{figure}

To answer this research question, we use the ML models with the best configuration of test history size, $H3$ for the \textit{Cisco} and \textit{ABB} datasets and $H2$ for the \textit{Google} dataset. We compare the fault-detection performance of the four ML models and two heuristics in terms of APFD, relative to the time budget available for running prioritized test suites. The results are shown in Figure~\ref{fig_image 13}. Columns $B1$-$B5$ correspond to five different values of the time budget, starting from 20\% of the average overall time required to run a test suite, with increments of 20\%. 

The results indicate that the $LRN$ approach and $ROCKET$ approach achieve similar performance on average. Both approaches perform better for longer time budgets, with $LRN$ having a slightly higher APFD for longer time budgets compared to $ROCKET$, and $ROCKET$ a slightly higher APFD for shorter time budgets compared to $LRN$ for some datasets, e.g. \textit{Cisco and ABB}. It is expected that longer time budget enables higher fault-detection, as there is more time available for testing, more test cases can be executed and more faults detected. $GBDT$ comes as the next best-performing approach, followed by $SVM$, on all datasets except \textit{Cisco}. For this particular dataset, $SVM$ slightly outperforms $GBDT$. $ANN$ approach has the worst fault-detection performance for short time budgets out of all ML-based approaches. Its performance improves for larger time budgets. Furthermore, $Random$ has the absolute worst fault-detection performance for short time budgets out of all evaluated approaches, while it fault-detection effectiveness improves for larger time budgets.

\subsection{RQ3: Time Effectiveness}
Time effectiveness is measured using four metrics: TDFF, TDLF, TRAIN, and PART. 
We report all the metrics in terms of the percentage of the time budget of a CI cycle. 

In terms of the time to detect the first fault (TDFF), $LRN$ has the best performance, which is comparable to $SVM$. The next best performing approach is $ROCKET$, followed by $GDBT$. The $ANN$ model has the worst ability to detect faults early out of all ML based approaches. However, $Random$ has the worst performance out of all evaluated approaches.

In terms of the time to detect the last fault (TDLF), $LRN$ shows to be a superior approach, followed by $SVM$ and $ROCKET$ which have comparable performance. The next best-performing approach is $GBDT$, followed by $ANN$. $Random$ shows the worst performance. 

In terms of the ML model training time (TRAIN), $LRN$ performs the best, followed by the $GBDT$ approach. $SVM$ is the third best performing approach in terms of training time, followed by $ANN$.  
 
In terms of the total running time of the prioritization algorithm (PART), $Random$ has the best performance. This is expected, since it uses a basic random test selection which is computationally cheap. The results further show that all ML approaches outperform $ROCKET$, which is expected. The $ANN$ approach has the best performance out of all ML approaches. $GBDT$ is the next best-performing approach, followed by $LRN$ and $SVM$. Average TRAIN and PART times for the three datasets \textit{Cisco}, \textit{ABB}, and \textit{Google} are shown in Table~\ref{tab2}.

\begin{table}[h]
\centering
\renewcommand{\arraystretch}{0.9}
\caption{Time metrics: average TRAIN and PART across \textit{Cisco}, \textit{ABB}, and \textit{Google} datasets.}
\scalebox{0.85}{
\begin{tabular}{c | c  c | c c | c c}
\toprule
 & \multicolumn{2}{c|}{Cisco} & \multicolumn{2}{c|}{ABB} & \multicolumn{2}{c}{Google}\\ 
\cmidrule(lr){2-3} \cmidrule(lr){4-5} \cmidrule(lr){6-7}
 
 & TRAIN [s] & PART [s] & TRAIN [s] & PART [s] & TRAIN [s] & PART [s] \\
\hline 
\textit{LRN} & 25 & 2 & 60 & 3 & 155 & 17\\
\textit{SVM} & 40 & 2.25 & 95 & 3 & 190 & 18\\
\textit{GBDT} & 35 & 1.5 & 90 & 2 & 175 & 15\\
\textit{ANN} & 50 & 1.35 & 105 & 1.9 & 199 & 10\\
\textit{ROCKET} & - & 65 & - & 125 & - & 3050\\
\textit{Random} & - & 1.25 & - & 1.8 & - & 9\\
\bottomrule
\label{tab2}
\end{tabular}}
\end{table}

\section{Discussion and Conclusion}
Test prioritization in continuous integration has the potential to improve the effectiveness and speed of fault detection. Machine learning has recently been proposed as an efficient approach for improving the scalability of test prioritization. Motivated by these findings, we set out to understand the relative fault-prediction performance of selected ML approaches for test case prioritization in continuous integration. We specifically focus on two parameters of continuous integration: test history size used for training ML models for test prioritization, and the size of time budget available for CI cycles. 

We selected four ML approaches that have shown good performance in test case prioritization in the literature (support vector machines, gradient boosting decision trees, neural networks, and LambdaRank with neural network) and designed a systematic experimental study comparing the four ML approaches one against the other and against the two heuristics for test prioritization. We compared the approaches in terms of time-effectiveness and fault-prediction effectiveness of prioritized test suites, answering three research questions. 
Our results show that the length of test execution history used for learning to prioritize test cases impacts the fault-prediction performance of ML approaches. For these datasets, our findings indicate that the optimal size of test history used for learning to prioritize is from 40\% to 60\%. When comparing different ML models for test prioritization, we observed that the performance of $ANN$ was the least sensitive to using older test history. At the same time, the $ANN$ approach showed the worst performance for short test history among all other evaluated ML approaches. Next, our results show that in terms of fault-prediction effectiveness for a given time budget, the best performing approach for a short time budget in terms of the APFD metric is $LRN$, while $ANN$ showed to have the worst fault-detection performance for a short time budget compared to the other evaluated ML-based approaches. Finally, in terms of time-effectiveness (time to detect the first and the last fault), the best performing approach is $LRN$. In terms of ranking time, the best performing ML approach is $ANN$.

\section*{References}

\bibliography{mybibfile}

\end{document}